\documentclass[a4paper,10pt]{article}

\usepackage[utf8x]{inputenc}

\usepackage{amsmath,amssymb,bm}
\usepackage{sidecap}
\usepackage{times}
\usepackage{braket}
\usepackage{graphicx,epsf}
\usepackage{float}
\usepackage[left=2.5cm,top=2.5cm,right=2.5cm,bottom=3cm,nohead,nofoot]{geometry}

\usepackage{xcolor} 

\date{}              

\begin{document}

\begin{centering}
\LARGE\textbf{Radiated fields by polygonal core-shell nanowires}
       
\vspace{12pt}      
\normalsize\textbf{Miguel Urbaneja Torres$^{\ast}$, Anna Sitek$^{\ast,\ddagger}$,   Vidar Gudmundsson$^{\P}$, and Andrei Manolescu$^{\ast}$}

\vspace{0pt}  
\normalsize\textit{$^{\ast}$School of Science and Engineering, Reykjavik University, Menntavegur 1, IS-101 Reykjavik, Iceland}\\
\normalsize\textit{$^{\ddagger}$Department of Theoretical Physics, Faculty of Fundamental Problems of Technology, Wroclaw University of Science and Technology, 50-370 Wroclaw, Poland}\\
\normalsize\textit{$^{\P}$Science Institute, University of Iceland, Dunhaga 3, IS-107 Reykjavik, Iceland}\\
\normalsize\textit{e-mail: miguelt16@ru.is}\\
\end{centering}
\vspace{6pt}

\noindent
\textbf{ABSTRACT}\\
We calculate the electromagnetic field radiated by 
tubular nanowires with prismatic geometry and infinite length. The
polygonal geometry has implications on the electronic localization;
the lowest energy states are localized at the edges of the prism and
are separated by a considerable energy gap from the states localized
on the facets.  This localization can be controlled with external
electric or magnetic fields.  In particular, by applying a magnetic field transverse
to the wire the states may become localized on the lateral regions of the
shell, relatively to the direction of the field, leading to channels of
opposite currents. Because of the prismatic geometry of the nanowire the
current distribution, and hence the radiated electromagnetic field, have
an anisotropic structure, which can be modified by the external fields.
In this work we study hexagonal, square and triangular nanowires.

\noindent


\vspace{12pt}
\noindent
\textbf{1. INTRODUCTION}
\vspace{3pt}

\noindent

Core-shell nanowires made of semiconductor materials, with diameters between 
a few tens and a few hundred nm, have recently
attracted great interest as a result of their rich electronic properties,
related both to transport and optics. Typically, such nanowires are
fabricated by the bottom-up method and have polygonal cross sections, most
often hexagonal \cite{Blomers13,Jadczak14}. Still, other
polygonal shapes are possible, like square \cite{Fan06} or triangular
\cite{Qian05,Heurlin15,Yuan15}. Interestingly, in such
structures the shell can be conductive and the core can be insulating,
such that one obtains a tubular conductor with a prismatic geometry.
In this geometry the electrons with low energies, situated within
the shell, tend to be localized along the prism edges, as the corners
of the cross section act like local quantum wells with a binding effect
\cite{Bertoni11a,Royo13,Fickenscher13}.  
In principle, for a polygon with $N$ corners, this group consist of $2N$
states, where the factor 2 accounts for the spin.  For thin shells
and sharp angles these states are nearly degenerate. Next on the energy
scale there is another group of $2N$ states localized on the prism facets.
Depending on the geometric parameters of the shell the two groups of
states can be separated by a remarkably large energy interval, possibly of
tens of meV, or even more for the triangular case \cite{Sitek15,Sitek16}.

In this work we study theoretically the electromagnetic fields radiated
by such prismatic shells. We obtain numerically the quantum mechanical states
and calculate the current along the nanowires considering a time-dependant
harmonic voltage bias. We discuss the implication of the corner
states on the radiated field for the three different
geometries. We also study the controllability of the filed distribution
with electric and magnetic fields external to the nanowire, which break
the spatial symmetry of the charge and current distributions within 
the shell.

\vspace{12pt}
\noindent
\textbf{2. THE MODEL}
\vspace{3pt}

\noindent
We focus on a system of non-interacting electrons confined in an infinite
nanowire with polygonal cross section. Our approach to model the polygonal
cross section begins with a circular ring, situated in the 
plane $(x,y)$,  described by
discretized polar coordinates \cite{Daday11}, on which
we superimpose polygonal constraints and retain the points situated
within the resulting shell. We assume free particle scattering
along the longitudinal direction $z$. The Hamiltonian of the system can be
expressed as:
\begin{equation}
\label{ham}
 H = \frac{(-i\hbar \nabla + e\bm{\mathrm{A}})^2}{2m_{\rm eff}} 
-e\bm{\mathrm{E}}\cdot\!\bm{r}\ 
-g_{\rm eff}\ \mu_{\rm B}\ {\bf \sigma}\cdot{\bf B}\ .
\end{equation}
We denote by ${\bf E}=(E_x,E_y,0)$ and ${\bf B}=(B_x,B_y,0)$
the transverse electric and magnetic fields, respectively, which
can be chosen at any angle relatively to the shell edges, and by
$\bm{\mathrm{A}}$ the vector potential associated with the magnetic
field. Also, ${\bm r}=(x,y,z)$ is the position vector within the shell volume, 
$e$ is the electron charge, $m_{\rm eff}$ and $g_{\rm eff}$ are the effective electron
mass and g-factor in the shell material, $\mu_{\rm B}$ is Bohr's magneton,
and  ${\bf \bm{\sigma}}=(\sigma_x,\sigma_y,\sigma_z)$ stands for the spin
related Pauli matrices.

We calculate the eigenstates of the Hamiltonian (\ref{ham}) numerically,
in two steps: first for ${\bf B} = 0$, to obtain the eigenvectors of the
transverse motion in the position representation, $\ket{a}$, where $a
=1,2,3...$, on a lattice with 6000-10000 points, depending on the geometry.
Then we use the first $2N$ modes, $N$ being the number of shell corners,
we form the basis set $\ket{aks}$, where $k$ is the wave vector corresponding to the
longitudinal motion, and $s=\pm 1$
is the spin label, and we diagonalize the total Hamiltonian for ${\bf B}\neq
0$, for a discretized series of $k$ values, to obtain its eigenvalues
$E_{mks}$ ($m=1,2,3,...$), and its eigenvectors $\ket{mks}$ expanded
in the basis $\ket{aks}$.  

This procedure gives us the states with the lowest energies, localized
along the edges or of the prismatic shells.  Using them we compute 
charge density $\rho$ and the current density ${\bm J}$ inside the shell as:
\begin{equation}
\label{expected_density}
\rho(\bm{r}) = e\sum_{mks} \! {\cal F} \left(\frac{E_{mks}-\mu}{k_{B}T}\right) 
\left[|\! \braket{\bm{r}|mks}\!|^2 - en_d  \right]\!, \
\bm{J}(\bm{r}) = \sum_{mks} \! {\cal F} \left(\frac{E_{mks}-\mu}{k_{B}T}\right) 
\braket{mks|\bm{j}(\bm{r}-\bm{r}_0)|mks}\!,
\end{equation}
where ${\cal F}(u) = 1/[\mathrm{exp}(u)+1]$
is the Fermi function with $u=(E_{mks}-\mu)/k_{B}T$, 
$\mu$ stands for the chemical potential,
$T$ the temperature, and $k_B$ Boltzmann's constant.  
The second term of the charge density represents the background of 
ionized donors of density $n_d$.
The operator $\bm{j}(\bm{r},\bm{r}_{0}) = e[\delta(\bm{r}-\bm{r}_0)\bm{v}
+ \bm{v}\delta(\bm{r}-\bm{r}_0)]/2$ describes
the contribution at spatial point $\bm{r}$ from
an electron situated at $\bm{r}_{0}$ which
moves with the velocity described by the operator $\bm{v}(\bm{r}_0) =
i[H,\bm{r}_0]/\hbar$. 

Obviously, in equilibrium, i.e. when no longitudinal voltage is applied
on the nanowire, the total current is zero; the current corresponding
to the electrons moving with positive velocity compensates the current
of those moving with negative velocity in the $z$ direction. In
order to generate a non-zero total current we simulate a voltage bias by creating
an imbalance between the states with positive and negative velocity,
i.e. with $\partial E_{mks}/ \partial k > 0$, and $\partial E_{mks}/
\partial k<0$, respectively \cite{Datta}.  To implement it we consider two
different chemical potentials, $\mu_{+}$ and $\mu_{-}$, associated with
positive and negative velocities, variable in time in a harmonic
manner, $\mu_{\pm} = \mu \pm V \sin(\omega t)$, i.e. with frequency $\omega$ 
and amplitude $2V$, where the static chemical 
potential $\mu$ is determined by the carrier density at equilibrium.
Using Eqs.\ (\ref{expected_density}) we compute the 
time dependent charge and current densities in the tubular shell, 
and then we calculate the radiated scalar and vector potentials outside
the nanowire, as


\begin{equation}
\label{potential}
\Phi_{\bm rad}(\bm{r},t) = \frac{1}{4 \pi \varepsilon _{0}} \int \frac{\rho (\bm{r'},t)}{|\bm{r}-\bm{r'}|} \mathrm{d}{\bm r'} \ , \hspace{1cm} 
\bm{A}_{\bm rad}(\bm{r},t) = \frac{\mu _0}{4 \pi} \int \frac{\bm{J} (\bm{r'},t)}{|\bm{r}-\bm{r'}|} \mathrm{d}{\bm r'} \ , 
\end{equation}
where the integration is performed within the shell domain, and $\varepsilon_0,
\mu_0$ are the vacuum constants.
Finally we obtain the radiated electric and magnetic fields using the Maxwell equations:

\begin{equation}
\label{field}
\bm{E}_{\bm rad} = - \nabla \Phi_{\bm rad} - \frac{\partial \bm{A}_{\bm rad}}{\partial t} \ , \hspace{1cm} 
\bm{B}_{\bm rad} = \nabla \times \bm{A}_{\bm rad} \ .
\end{equation}

In our present approach we neglect the retardation effects, since
the speed of light can be considered infinite at the nanometric scale.
In addition we consider an arbitrary (unspecified) frequency $\omega$, but in principle 
sufficiently low to prevent strong damping. Our main goal is to show
qualitatively the fingerprint of the tubular prismatic geometry on the spatial
structure of the radiated field, and in particular the combined effect 
of localization and external fields.

\vspace{12pt}
\noindent
\textbf{3. RESULTS}
\vspace{3pt}

In this section we show qualitatively some examples of radiated field configurations for
the three polygonal geometries: hexagonal, square, and triangular. For simplicity we  represent
the radiated field by the corresponding magnetic component. We use
InAs bulk parameters for the shell : $m_{\rm eff}=0.023m_e$ and $g_{\rm eff}=-14.9$.
In all cases the external radius of the polygonal shell is $R_{\rm ext}=50$
nm and the thickness of the facets is $t=10$ nm.  Also, the amplitude of
the AC voltage bias used is fixed to $2V = 5$ meV.

\begin{figure} [H]
\centering
\includegraphics[scale=0.80]{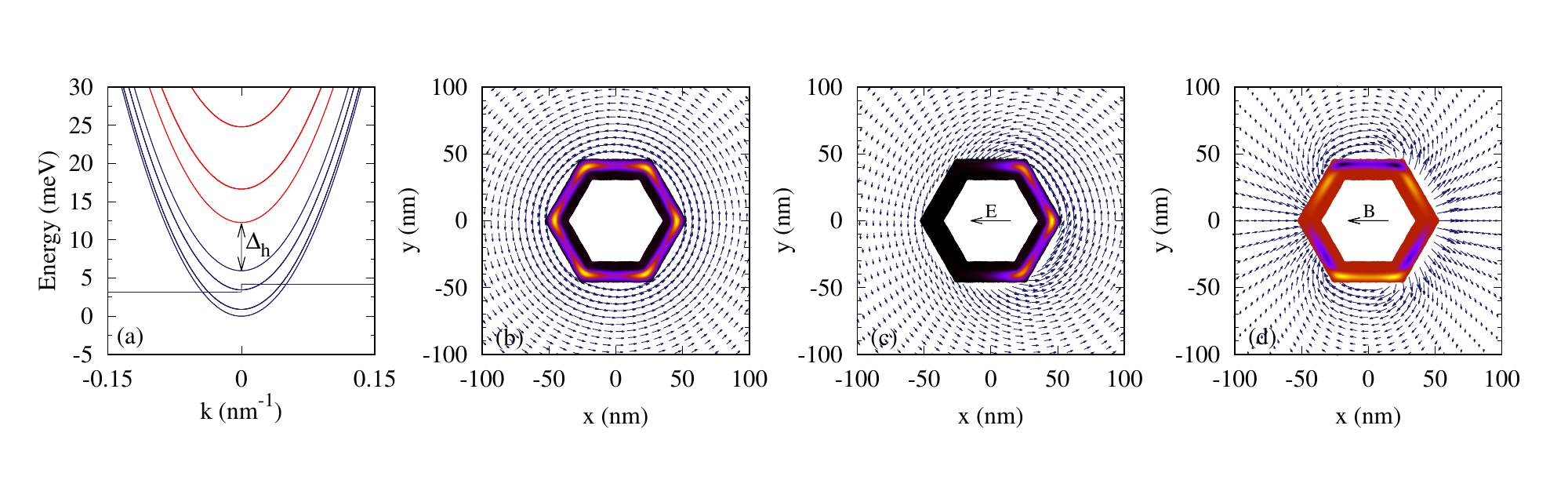}

\vspace{-9mm}
\caption{\textit{Hexagonal shell. (a) Energy spectrum in the absence of
external fields. The blue curves correspond to the corner-localized states
and the red ones to the side-localized states. The energy separation
between them is $\Delta_{h}=6.5$ meV. The horizontal lines indicate the
biased chemical potential at maximum amplitude $2V=5$ meV. (b) The current 
distribution and the radiated field in the absence of external fields. The current 
distribution inside the shell is shown on a color scale, with the yellow 
and purple colors corresponding to electrons propagating in the 
positive and negative $z$ direction, respectively, and with black color corresponding 
to regions with vanishing electron density.  The radiated magnetic field 
outside the nanowire is represented with arrows.  
(c) The current distribution and the radiated magnetic field in the presence
of an external electric field $E=10 \ {\rm \mu V/nm}$. (d) The current distribution and the radiated 
magnetic field
in the presence of an external magnetic field ${\bm B=1}$ T. 
}}
\label{Fig_hex}
\vspace{-1mm}
\centering
 \includegraphics[scale=0.8]{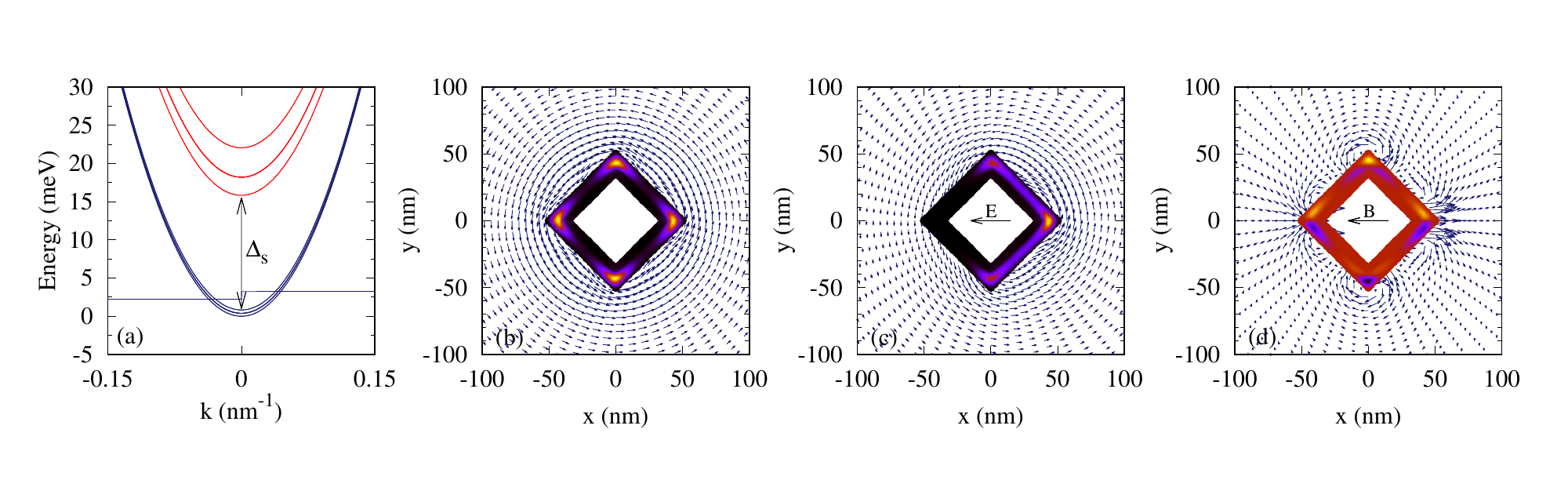}

\vspace{-9mm}
\caption{\textit{Similar to Fig.\ \ref{Fig_hex}, but for a square shell.  In this case $\Delta_{s}=14.6$ meV}.} 
\label{Fig_squ}
\vspace{-1mm}
\centering
 \includegraphics[scale=0.8]{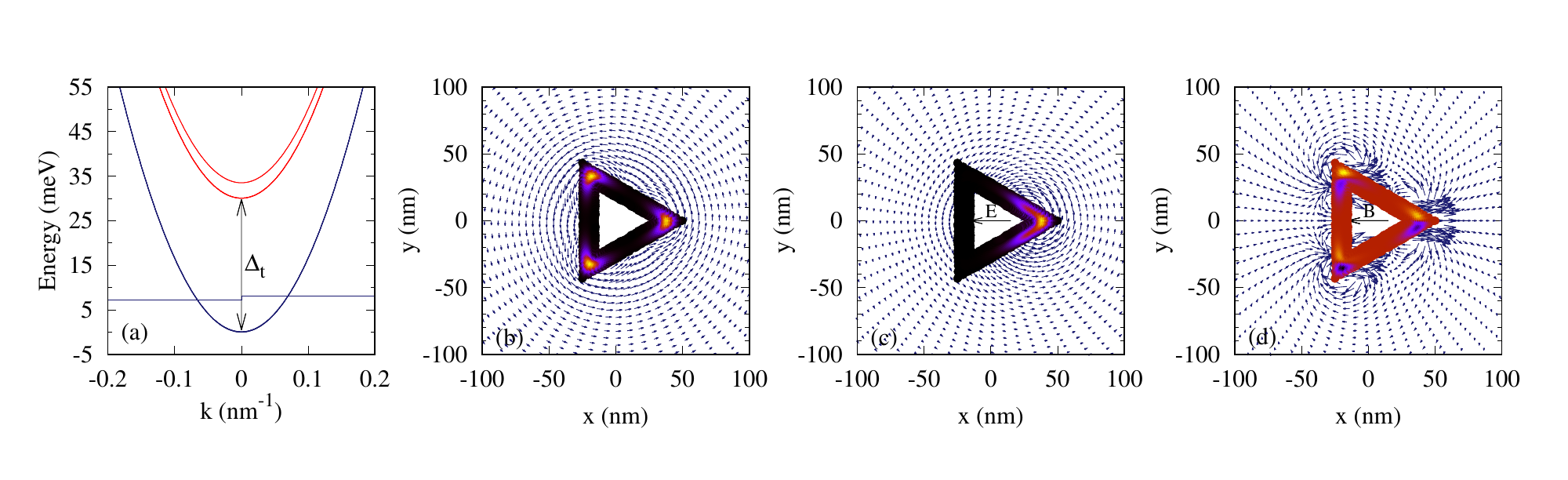}

\vspace{-9mm}
\caption{\textit{Similar to Figs.\ \ref{Fig_hex} and \ref{Fig_squ}, but now for a triangular shell, with 
$\Delta_{t}=27.1$ meV}.} 
\label{Fig_tri}
\end{figure}

In Figures \ref{Fig_hex}, \ref{Fig_squ}, and \ref{Fig_tri} we first
show, for each geometry, the energy spectra for the symmetric shells.
In Panels (a) we indicate with blue lines the corner states,
with red lines the side states, and the energy interval between them
with $\Delta$.  We assume that the chemical potential is sufficiently low
(or, equivalently, the electron density is low), such that only the corner
states are populated. With this condition the anisotropy related to the
internal structure of the nanowire is maximal.
As mentioned before, we obtain twelve corner states for the hexagonal,
eight for the square, and six for the triangular geometries.  Because of
the spin and rotational symmetries these states can be two or four-fold
degenerate, such that multiple blue (and also red) lines overlap 
in Panels (a) of each figure.

Next, we show the current distributions and the radiated fields for
three situations:  first the nanowires with the polygonal symmetric
distribution of electrons, and then in the presence of the transverse
electric or magnetic fields.
If the polygonal shell is perfectly symmetric the electrons are equally
distributed between the corners. In this case, in the presence of a
chemical potential bias, currents are running in parallel along the prism
edges, as shown in Panels (b).  The radiated electromagnetic field from
each edge is similar to the one created by a line of current.
The total field is thus a superposition of the $N$ edge currents, with
the spatial symmetry of the prismatic shell.

An external transverse (static) electric field breaks this spatial symmetry
of the charge distribution by pushing the electrons laterally, and allowing
for the controllability of the localization.  This perturbation leads to 
a different structure of the radiated field in the vicinity of the nanowire,
as shown in Panels (c). We notice that the anisotropy of the radiated field 
increases for the square and furthermore for the triangular shell.
The Panels (b) and (c) correspond to the instant in time when the
voltage bias is maximized, i.e. $\mu_{+}-\mu_{-}=2V$.

A much more interesting situation occurs when a magnetic field is applied 
perpendicular to the nanowire.  In Panels (d) we show the case when it is 
applied along a certain symmetry axis of the prismatic shell.  In this situation
the corner localization coexists with the localization induced by the magnetic field. 
The magnetic localization is determined by the motion of the electrons along the 
nanowire and can lead to the formation of local Landau states in the regions where
the radial component of the magnetic field is locally constant, and of 
snaking states in the regions where it changes sign.  The density of electrons
tends to increase around the snaking states \cite{Manolescu13}.  
Consequently, we obtain channels of current
traveling in opposite directions. These channels can also be considered as large 
loops of current along the nanowire length driven by the Lorentz force. 
To illustrate this fact we selected the instant in time when the chemical potential
bias is zero and there is no net current through the nanowire, 
each current channel being compensated by its pair. 
Remarkably, the paired channels can be 
situated on opposite sides of the shell, but also within the same corner. 
In the hexagonal case the external magnetic field induces three
loops: one along the facets parallel to the magnetic field, and two
more in the other facets, Fig.\ \ref{Fig_hex}(d). In the case shown for
the square geometry we obtain current channels only in the 
corner areas: snaking states in the corners lateral to the direction of the 
field and local Landau states in the other two corners. 
%
%
In the example shown for the triangular shell, Fig.\ \ref{Fig_tri}(d), where
the binding effect is the strongest, the electrons form three loops: two of
them within the facet perpendicular to the magnetic field
and one in the opposite corner, all of Landau type.


Thus, in the presence of a transverse magnetic field, the radiated field
can be much richer in structure than in the previous cases. 
The magnetic field radiated can be highly anisotropic near the shell,
especially for the square and triangular geometries, and resembles the
field created by a magnetic dipole.

\vspace{12pt}
\noindent
\textbf{4. CONCLUSIONS}
\vspace{3pt}

\noindent
We discussed the magnetic component of the electromagnetic field radiated
by polygonal core-shell nanowires exposed to transverse
electric or magnetic fields, which allow to control the localization
of electrons. We showed that the internal geometry of the nanowire has
implications on the configuration of the radiated field.  In particular,
an external transverse magnetic field induces
longitudinal channels of electrons traveling in opposite directions,
which may lead to a highly anisotropic radiated field.


\vspace{12pt}
\noindent
\textbf{ACKNOWLEDGEMENTS}
\vspace{3pt}

\noindent
This work was financed by the Icelandic Research Fund. 
We are thankful to Sigurdur Erlingsson 
for discussions.

\bibliographystyle{unsrt}

\begin{thebibliography}{14}

\bibitem{Blomers13}
C.~Bl{\"o}mers, T.~Rieger, P.~Zellekens, F.~Haas, M.~I. Lepsa, H.~Hardtdegen,
  {\"O}.~G{\"u}l, N.~Demarina, D.~Gr{\"u}tzmacher, H.~L{\"u}th, and
  Th.~Sch{\"a}pers,
\newblock {\em Realization of nanoscaled tubular conductors by means of GaAs/InAs
  core/shell nanowires},
\newblock Nanotechnology \textbf{24}, 035203 (2013).

\bibitem{Jadczak14}
J.~Jadczak, P.~Plochocka, A.~Mitioglu, I.~Breslavetz, M.~Royo, A.~Bertoni,
  G.~Goldoni, T.~Smolenski, P.~Kossacki, A.~Kretinin, H.~Shtrikman, and
  D.~K. Maude,
\newblock {\em Unintentional high-density p-type modulation doping of a GaAs/AlAs
  core–multishell nanowire},
\newblock Nano Letters \textbf{14}, 2807 (2014).

\bibitem{Fan06}
H.~Fan, M.~Knez, R.~Scholz, K.~Nielsch, E.~Pippel, D.~Hesse, U.~G{\"o}sele, and M.~Zacharias,
\newblock {\em Single-crystalline mgal 2 o 4 spinel nanotubes using a reactive and
  removable mgo nanowire template},
\newblock Nanotechnology \textbf{17}, 5157 (2006).

\bibitem{Qian05}
F.~Qian, S.~Grade{\v{c}}ak, Y.~Li, Ch.-Y.~Wen, and Ch.~M.~Lieber,
\newblock {\em Core/multishell nanowire heterostructures as multicolor,
  high-efficiency light-emitting diodes},
\newblock Nano Letters \textbf{5}, 2287 (2005).

\bibitem{Heurlin15}
M.~Heurlin, T.~Stankevi{\v{c}}, S.~Mickevi{\v{c}}ius, S.~Yngman, D.~Lindgren, 
A.~Mikkelsen, R.~Feidenhans’l, M.~T.~Borgst{\"o}m, and L.~Samuelson,
\newblock {\em Structural properties of wurtzite InP–InGaAs nanowire core–shell
  heterostructures},
\newblock Nano Letters \textbf{15}, 2462 (2015).

\bibitem{Yuan15}
X.~Yuan, Ph.~Caroff, F.~Wang, Y.~Guo, Y.~Wang, H.~E.~Jackson, L.~M.~Smith, 
H.~H.~Tan, and Ch.~Jagadish,
\newblock {\em Antimony induced {112}a faceted triangular GaAs$_{1−x}$Sb$_{x}$/InP
  core/shell nanowires and their enhanced optical quality},
\newblock Adv. Funct. Mater. \textbf{25}, 5300 (2015).

\bibitem{Bertoni11a}
A.~Bertoni, M.~Royo, F.~Mahawish, and G.~Goldoni,
\newblock {\em Electron and hole gas in modulation-doped
  GaAs/Al$_{\mathrm{1-x}}$Ga$_\mathrm{x}$As radial heterojunctions},
\newblock Phys. Rev. B \textbf{84}, 205323 (2011).

\bibitem{Royo13}
M.~Royo, A.~Bertoni, and G.~Goldoni,
\newblock {\em Landau levels, edge states, and magnetoconductance in GaAs/AlGaAs
  core-shell nanowires},
\newblock Phys. Rev. B \textbf{87}, 115316 (2013).

\bibitem{Fickenscher13}
M.~Fickenscher, T.~Shi, H.~E.~Jackson, L.~M.~Smith, J.~M.~Yarrison-Rice, 
Ch.~Zheng, P.~Miller, J.~Etheridge, B.~M.~Wong, Q.~Gao, S.~Deshpande, H.~H.~Tan, 
and Ch.~Jagadish,
\newblock {\em Optical, structural, and numerical investigations of GaAs/AlGaAs
  core–multishell nanowire quantum well tubes},
\newblock Nano Letters \textbf{13}, 1016 (2013).

\bibitem{Sitek15}
A.~Sitek, L.~Serra, V.~Gudmundsson, and A.~ Manolescu,
\newblock {\em Electron localization and optical absorption of polygonal quantum
  rings},
\newblock Phys. Rev. B \textbf{91}, 235429 (2015).

\bibitem{Sitek16}
A.~Sitek, G.~Thorgilsson, V.~Gudmundsson, and A.~Manolescu,
\newblock {\em Multi-domain electromagnetic absorption of triangular quantum rings},
\newblock Nanotechnology \textbf{27}, 225202 (2016).

\bibitem{Daday11}
C.~Daday, A.~Manolescu, D.~C. Marinescu, and V.~Gudmundsson,
\newblock {\em Electronic charge and spin density distribution in a quantum ring
  with spin-orbit and coulomb interactions},
\newblock Phys. Rev. B \textbf{84}, 115311 (2011).

\bibitem{Datta}
S.~Datta,
\newblock {\em Electronic Transport in Mesoscopic Systems},
\newblock Cambridge University Press (1997).

\bibitem{Manolescu13}
A.~Manolescu, T.~O.~Rosdahl, S.~I.~Erlingsson, L.~Serra and V.~Gudmundsson,
\newblock {\em  Snaking states on a cylindrical surface in a perpendicular magnetic field},
\newblock Eur. Phys. J. B. \textbf{86}, 445 (2013).


\end{thebibliography}

\end{document}